\documentclass[twocolumn,superscriptaddress, floatfix, amsfonts, aps]{revtex4-1} 
\usepackage{amsmath}
\usepackage{bm}
\usepackage{graphicx}
\usepackage{subfigure}
\usepackage{float}
\usepackage{mathrsfs}
\usepackage{footmisc}
\usepackage{multirow}
\usepackage{graphicx}
\usepackage{soul,xcolor}
\usepackage{xcolor, ulem, color, framed}
\usepackage{amssymb}
\usepackage{tikz}

\usepackage[colorlinks, linkcolor=red,anchorcolor=green,citecolor=blue]{hyperref}

\usepackage[colorlinks, linkcolor=red,anchorcolor=green,citecolor=blue]{hyperref}

\graphicspath{{./figures/}}

\begin{document}

 \title{Charmonia Production in Hot QCD Matter and Electromagnetic Fields}

 \author{Shuhan Zheng}
 \affiliation{Department of Physics, Tianjin University, Tianjin 300354, China}
 \affiliation{Department of Physics, Tsinghua University, Beijing 100084, China}
  \author{Jiamin Liu}
  \affiliation{Department of Physics, Tianjin University, Tianjin 300354, China}
  \author{Shiqi Zheng}
  \affiliation{School of Engineering, Brown University, Providence, Rhode Island 02912-1843, USA}
    \author{Jiaxing Zhao}
\affiliation{Institute for Theoretical Physics, Johann Wolfgang Goethe Universit\"at, Frankfurt am Main, Germany}
\affiliation{
Helmholtz Research Academy Hessen for FAIR (HFHF), GSI Helmholtz Center for Heavy Ion Research. Campus Frankfurt, 60438 Frankfurt, Germany}
 \author{Baoyi Chen}\email{baoyi.chen@tju.edu.cn}
\affiliation{Department of Physics, Tianjin University, Tianjin 300354, China}

\begin{abstract}
Both hot QCD matter and extremely strong electromagnetic fields are generated in relativistic heavy-ion collisions. We employ the transport model and the equivalent photon approximation (EPA) to study charmonium hadroproduction and photoproduction in nucleus-nucleus collisions, respectively. In photoproduction, quasi-real photons may interact with the whole nucleus or individual nucleons, which is called the coherent and incoherent processes, respectively. The typical momentum of charmonium produced in two processes is located in $p_T\lesssim 1/R_A$ and $p_T\lesssim 1/R_N$, where $R_A$ and $R_N$ are the radii of nucleus and the nucleon. Both kinds of photoproduction and also hadroproduction are considered to calculate 
charmonium production in different transverse momentum bins, rapidity bins, and collision centralities, incorporating modifications from hot QCD matter and initial cold nuclear matter effects. Our calculations explain experimental data about charmonium nuclear modification factors and the production cross-section in ultra-peripheral collisions. Charmonium nuclear modification is far above the unit at extremely low $p_T$ ($p_T < 0.1$ GeV/c) in peripheral collisions with centrality 70-90\%, attributed to coherent photoproduction.

\end{abstract}
\date{\today}
\maketitle

\section{Introduction}

Hot QCD matter is one of the crucial topics in high-energy nuclear physics, which can be created in experiments through relativistic heavy-ion collisions~\cite{STAR:2005gfr,PHENIX:2004vcz}. In collisions, abundant partons are produced, forming a deconfined state known as Quark-Gluon Plasma (QGP)~\cite{Bazavov:2011nk}. Over the past decades, extensive studies have studied the signals and properties~\cite{Ollitrault:1992bk,Schafer:2009dj,PHENIX:2008uif,Heinz:2013th,Huang:2009ue,Qin:2015srf,Zhang:2003wk} of hot QCD matter. Charmonium, a tightly bound state of charm and anti-charm quarks with significant binding energy, undergoes dissociation in hot QCD matter due to color screening and parton inelastic scatterings~\cite{Matsui:1986dk,Satz:2005hx,Du:2017qkv,Yan:2006ve}, leading to yield suppression in nuclear collisions compared with the case of pp collisions. The degree of suppression of the quarkonium yield increases with the temperature of the medium ~\cite{Liu:2010ej,Chen:2019qzx}. This suppression of $J/\psi$ has been proposed as a clean probe of the deconfined matter created in the early stages of heavy-ion collisions~\cite{Matsui:1986dk}. 

In nuclear collisions at LHC energies, many charm-anticharm pairs are also produced~\cite{ALICE:2015vxz,ALICE:2018lyv}. These uncorrelated charm and anticharm quarks can recombine to form new charmonium states in the deconfined phase when the heavy quark potential is partially restored, labeled as regeneration~\cite{Andronic:2003zv,Greco:2003vf,Zhao:2017yan,Chen:2017duy,Blaizot:2018oev}. This process contributes to the final production of charmonium. These charm quarks are strongly coupled with the QGP, experiencing energy loss and carrying collective flows within the expanding bulk medium~\cite{Beraudo:2007ky,Zhao:2023nrz,He:2012df,Cao:2013ita}. Consequently, regenerated charmonium is primarily found in the intermediate and low transverse momentum regions ($p_T < 5$ GeV/c)~\cite{Du:2015wha,Chen:2018kfo,Pan:2023ouw}. At high $p_T$, the charmonium yield is dominated by primordial production~\cite{Zhou:2014kka,Du:2017qkv}.

When a nucleus moves at extremely high velocities, it also generates strong electromagnetic fields~\cite{Deng:2012pc}. These fields can influence the dynamics of quarks in the QGP, such as the Chiral Magnetic Effect~\cite{Fukushima:2008xe,Shi:2017cpu,Shi:2019wzi}, Chiral Vortical Effect~\cite{Jiang:2015cva}, and spin polarization~\cite{Huang:2020dtn,Jian-Hua:2023cna,Sheng:2024pbw,Liu:2024hii,Liang:2004ph}. Magnetic fields can also dissociate charmonium states, contributing to the anisotropy of the charmonium momentum distribution~\cite{Guo:2015nsa}. This mechanism differs from the anisotropy of the charmonium momentum distribution resulting from the anisotropic expansion of the bulk medium.
Furthermore, the transverse electromagnetic fields generated by the fast-moving nucleus can produce vector mesons, called photoproduction~\cite{Klein:2020fmr}.
The electromagnetic fields are treated as quasi-real photons, which interact with the other nucleus to produce vector mesons~\cite{Klein:1999qj,ALICE:2015mzu,Cepila:2017nef} and dileptons~\cite{Zha:2018tlq,Li:2018wuv,Luo:2023syp}, known as the Equivalent Photon Approximation (EPA) method~\cite{Fermi:1924tc,vonWeizsacker:1934nji}. With different energies, quasi-real photons may interact with the whole nucleus or nucleons inside; the corresponding processes are called coherent and incoherent processes. The typical transverse momentum of photoproduced charmonium depends on the energy of quasi-real photons, which can be estimated with the uncertainty relation to be $p_T\lesssim 1/R_A$ and $p_T\lesssim 1/R_N$ in coherent and incoherent scatterings. $R_A$ and $R_N$ are the radii of the nucleus and nucleon, respectively.
Different from ultra-peripheral collisions, in peripheral and semi-central collisions, both photoproduction and hadroproduction contribute to the final yield of charmonium, with modifications from hot QCD matter~\cite{Shi:2017qep}.

This work studies charmonium production from hadroproduction and photoproduction from ultra-peripheral to central nuclear collisions at LHC and RHIC energies. Sections II and III introduce the theoretical frameworks for photoproduction and hadroproduction, respectively. Both coherent and incoherent processes are considered. Section IV presents the initial conditions and the cold nuclear matter effects used in hadroproduction. Section V presents numerical results for the charmonium production cross-section in ultra-peripheral collisions, compared with the experimental data. This section also comprehensively analyzes charmonium nuclear modification factors in different $p_T$ and rapidities. Section VI provides a final summary.

\section{Photoproduction framework}

Strong electromagnetic fields are generated by the protons in heavy-ion collisions, with Lorentz contraction in the longitudinal direction. Assuming a Woods-Saxon distribution of protons in the nucleus, the normalized spatial density of protons, $\rho^{\rm norm}(r)$, in the nucleus is parametrized as,
\begin{eqnarray}
\rho^{\rm norm}(r)&=&{\rho_0 \over 1+\exp({r-r_0\over a}) },
\label{eq.ws}
\end{eqnarray}
with the parameters $r_0 = 6.624$ fm and $a = 0.549$ fm for the nucleus Pb. The normalization factor $\rho_0$ is determined by $\int d^3{\bf r} \, \rho^{\rm norm}(r) = 1$. The corresponding electromagnetic form factors, which characterize the electric charge density of the nucleus in momentum space, are obtained through a Fourier transform,
\begin{eqnarray}
F({\bf  q})=\int \rho^{\rm norm}(r)e^{i {\bf q}\cdot {\bf r}}d^3{\bf r}.
\label{eq.form}
\end{eqnarray}
According to the equivalent photon approximation, the density of quasi-real photons is connected with the nuclear electric charge densities via the relation~\cite{Shi:2017qep,Zha:2017jch} 
\begin{eqnarray}
&&n(\omega,{\bf x_T})=
\nonumber\\&&
{4Z^2\alpha \over \omega} \left|\int {d^2 {\bf k_T}\over (2\pi)^2} k_T {F(\sqrt{k_T^2+\omega^2/\gamma_L^2}) \over k_T^2+\omega^2/\gamma_L^2}e^{i {\bf x_T} \cdot {\bf k_T}}\right|^2,
\label{eq.photon}
\end{eqnarray} 
where $w$ is the photon energy, ${\bf x_T}$ and ${\bf k_T}$ represent the transverse coordinate and transverse momentum of the photon, respectively. Here, $Z = 82$ denotes the number of protons in the Pb nucleus, and $\alpha = 1/137$ is the fine structure constant. The parameter $\gamma_L$ represents the Lorentz factor associated with the motion of protons in the longitudinal direction. 

In heavy-ion collisions with impact parameter ${\bf b}$, the quasi-real photons from one nucleus (called the source nucleus) interact with the other nucleus (called target nucleus), generating charmonium based on the photon density given by Eq.(\ref{eq.photon}). In the ultra-peripheral collisions (UPC) without hadronic interactions, 
the cross-section for $J/\psi$ photoproduction from the coherent process is~\cite{Krauss:1997vr,Klein:2003vd},
\begin{align}
\label{eq-upccross}
\frac{d\sigma_{J/\psi}^{\rm UPC}}{dy}(y)=&\omega n_\gamma^{\rm UPC}(\omega)\sigma^{\rm coh}_{\gamma A\to J/\psi A}(\omega)\\ \nonumber 
&+
\omega' n_\gamma^{\rm UPC}(\omega')\sigma^{\rm coh}_{\gamma A\to J/\psi A}(\omega')
\end{align}
where $w = e^y m_{J/\psi}/2$ and $w' = e^{-y} m_{J/\psi}/2$~\cite{Klein:2020fmr} are the energies of photons emitted from the two nuclei, respectively. $m_{J/\psi}$ is the mass of $J/\psi$ and $y$ denotes the rapidity of the photoproduced $J/\psi$. The energy spectrum of quasi-real photons, $n_\gamma^{\rm UPC}(w) = dN_\gamma/dw$, describes photons that will interact coherently with the target nucleus. In UPC, the impact parameter is defined as being greater than twice the nuclear radius $b \ge 2R_A$. The photon energy spectrum is the integration of Eq.(\ref{eq.photon}),
\begin{align}
n_\gamma^{\rm UPC}(\omega)  = \int_{2R_A}^{\infty} \int_0^{2\pi} n(w,{\bf r}) rdrd\theta,
\label{eq-upcden}
\end{align}
where $r$ is the radius in the transverse plane $r=|{\bf x_T}|$. The other ingredient in Eq.(\ref{eq-upccross}) is the photon-nucleus coherent cross-section $\sigma^{\rm coh}_{\gamma A\rightarrow J/\psi A}$. The momentum dependence in the cross-section is introduced with the form factor,
\begin{align}
\sigma^{\rm coh}_{\gamma A\to J/ \psi A}=\frac{d\sigma^{\rm coh}_{\gamma A\to  J/ \psi A}}{dt}\bigg|_{t=0}\int_{t_{min}}^{\infty}dt|{F}(t)|^2.
\label{eq.sigmaA}
\end{align}
Here $t_{min}= [{m_{J/\psi}^2}/({4\omega\gamma_L})]^2$ is the minimal momentum squared needed to produce a vector meson. $t=p_T^2$ is the square of the $J/\psi$ transverse momentum. ${F}(t)$ is the form factor, given by Eq.(\ref{eq.form}). The photon-nuclear differential cross-section is derived with the photon-proton cross-section via optical theorem and Glauber calculations, parametrized as~\cite{Klusek-Gawenda:2015hja,Klein:1999qj}, 
\begin{align}
\frac{d\sigma^{\rm coh}_{\gamma A\to  J/\psi A}}{dt}|_{t=0}&=C^{2}\frac{\alpha\sigma_{tot}^{2}(J/\psi A)}{4f_{V}^{2}}, \\
\sigma_{tot}(J/\psi A)&=2\int d^2{\bf x_T}[1-e^{-\frac{1}{2}\sigma_{tot}(J/\psi p)T_A({\bf x_T})}], \\
\sigma_{tot}^{2}(J/\psi p)&=\frac{4f_{V}^{2}}{\alpha C^{2}}\frac{d\sigma(\gamma p\to  J/\psi p)}{dt}|_{t=0}. \label{eq.photo-p}
\end{align}
The value of vector meson coupling constant $f_V^2$ is taken as $f_V^2=10.4\times 4\pi$ for $J/\psi$~\cite{Shi:2017cpu}, according to the generalized vector dominance model (GVDM)~\cite{Pautz:1997eh}. The thickness function of the nucleus Pb is given by $T_A=\int dz \rho_N({\bf x_T},z)$, with a total nucleon number of 208. Therefore, $\rho_N(r)=208\rho^{\rm norm}(r)$ as specified in Eq.(\ref{eq.ws}).  $C=0.3$ is a correction factor~\cite{Hufner:1997jg,Zha:2017jch}. 
The photon-proton differential cross-section is parametrized with HERA data~\cite{hera-ref},
\begin{align}
\frac{d\sigma(\gamma p\to J/\psi p)}{dt}|_{t=0}=b_v X W_{\gamma p}^\epsilon\times\left[1-(\frac{m_{J/\psi}+m_p}{W_{\gamma p}})^2\right],
\end{align}
where $b_V=4.0 \text{ GeV}^{-2}$, $X=0.00406 \text{ $\mu$b}$ and $\epsilon =0.65$~\cite{Shi:2017qep}. $W_{\gamma p}= 2\sqrt{\omega E_p}$ is the center-of-mass energy of the photon and proton. The proton energy is half the collision energy $E_p=\sqrt{s_{NN}}/2$.

In peripheral and semi-central collisions, the electromagnetic fields and photon density differ with the impact parameter. With a larger value of the impact parameter, the density of quasi-real photons located in the target nucleus becomes smaller, which induces smaller photoproduction of vector mesons. Besides, the density of quasi-real photons $n(w, {\bf r})$ emitted by the source nucleus also varies in the area of the target nucleus. In coherent scattering process, photons interact with the whole target nucleus. Therefore, we employ the averaged density of quasi-real photons in the area of the target nucleus to calculate the coherent photoproduction, 
\begin{align}
n_\gamma(\omega|b)&=
{1 \over \pi R_A^2}\int_0^{R_A}rdr\int_0^{2\pi}d\phi n(\omega, r'),
\label{eq.photonaver}
\end{align}
where $r'=\sqrt{b^2+2br\cos\phi+r^2}$ and $b$ is the impact parameter. In Eq.(\ref{eq.photonaver}), the photon density is averaged over the area of $\pi R_A^2$ with a fixed impact parameter. 

In the forward and backward rapidity region, the contribution of incoherent photoproduction is not negligible for $J/\psi$ production~\cite{Liu:2023uni}, especially in the region $1/R_A<p_T<1/R_N$ which is around $0.1$ GeV/c.
To include this contribution, inspired by the Glauber model and GVDM, the incoherent cross section $d\sigma_{\gamma A\to J/\psi A^{\prime}}^{\rm inc} / dt$ can be expressed by the following formulas 
\begin{align}
\frac{d\sigma_{\gamma A\to J/\psi A^{\prime}}^{\rm inc} }{dt}  &=\frac{d\sigma_{\gamma p\to J/\psi p}^{\rm inc}}{dt} \notag \\
\times&\int T_{A}({\bf x_T})e^{-{1\over 2}\sigma_{J/\psi p}^\mathrm{in}T_A({\bf x_T})}d{\bf x_T}, \\
\frac{d\sigma_{\gamma p\to J/\psi p}^{\rm inc}}{dt}  &= \frac{d\sigma_{\gamma p\to J/\psi p}}{dt} |_{t=0} |F_{\rm hcs}(t)|^2, \\
\sigma_{J/\psi p}^\mathrm{in} &= \sigma_{tot}(J/\psi p)-\sigma_{tot}^{2}(J/\psi p)/(16\pi B_V),
\end{align}
where $B_V$ = 4.0 GeV$^2$ and $\sigma_{J/\psi p}^{\mathrm{in}}$ represents inelastic vector $J/\psi$-proton cross section~\cite{Guzey:2020pkq}. With regard to the incoherent photon-proton cross section $d\sigma_{\gamma p\to J/\psi p}^{\rm inc} / dt$, the dependence of transverse momentum $p_T$ is introduced via the homogeneous charged sphere (HCS) form factor $F_{\rm hcs}(t)$ in the proton, which can be expressed as
\begin{align}
F_{\rm hcs}(q^2) = \frac{3 j_1 (qR)}{qR},
\label{eq:formproton}
\end{align}
in which $q$ and $R=1$ fm (treated as the proton radius) signify the momentum and the radius of HCS, respectively. $j_1(qR)$ represents the spherical Bessel function.

Combing both coherent and incoherent cross-sections, 
the $p_T$ differential distribution of $J/\psi$ in peripheral and semi-central collisions can be easily extended as, 
\begin{align}
{dN_{J/\psi}\over 2\pi p_T dp_T dy} &={1\over \pi \Delta y}\int_{y_{\rm min}}^{y_{\rm max}} \Big[\omega n_\gamma(\omega|b){d\sigma^{\rm coh+inc}_{\gamma  A}(\omega,t)\over dt} \notag
\\ &+\omega^\prime n_\gamma(\omega^\prime|b){d\sigma^{\rm coh+inc}_{\gamma A}(\omega^\prime,t)\over dt}\Big]dy,
\label{eq.ptphotores}
\end{align}
where $\Delta y=y_{\rm max}-y_{\rm min}$ with $y_{\rm max}$ and $y_{\rm min}$ characterize the rapidity bins selected in experimental data. $d\sigma^{\rm coh+inc}_{\gamma A}/dt=d\sigma_{\gamma A\rightarrow J/\psi A}^{\rm coh}/dt+ d\sigma_{\gamma A\rightarrow J/\psi A'}^{\rm inc}/dt$.
The relation $dt=2p_Tdp_T$ is employed.

\section{Hadroproduction} 
In heavy-ion collisions, charmonium hadroproduction includes both prompt production and non-prompt production from B-hadron decay~\cite{Chen:2013wmr}. The prompt production consists of primordial production and regeneration. A Boltzmann-type transport model has been developed to study the evolution of prompt charmonium in phase space~\cite{Zhu:2004nw}. Defining the proper time $\tau = \sqrt{t^2 - z^2}$ and pseudorapidity $\eta = \frac{1}{2} \ln \left( \frac{t + z}{t - z} \right)$, the transport equation is written as, 
\begin{align}
&\left[\cosh(y-\eta)\frac\partial{\partial\tau}+\frac{\sinh(y-\eta)}\tau\frac\partial{\partial\eta}+\boldsymbol{v}_t\cdot\nabla_t\right]f_\psi \nonumber \\
&=-{\alpha_\psi}f_\psi+{\beta_\psi},
\end{align}
where $\alpha_\psi$ is the decay rate of quarkonium in the hot medium, induced by the combined effects of color screening and parton inelastic scatterings. $\beta_\psi$ represents the rate of recombination of charm and anti-charm quarks into charmonium states. In the dissociation reaction $g + J/\psi \leftrightarrow c + \bar{c}$, the decay rate $\alpha_\psi$ depends on both the density of thermal gluons and the inelastic cross section~\cite{Du:2017qkv},
\begin{align}
{\alpha_\psi}=\frac{1}{2E_T}\int\frac{d^3\bf{p}_g}{(2\pi)^3 2E_g}W_{g\psi}^{c\bar{c}}(s)f_g({\bf p}_g)\Theta(T(x)-T_c),
\end{align}
where ${\bf p}_g$ and $E_g$ denote the momentum and energy of the thermal gluon, respectively. $E_T=\sqrt{m_{\psi}^2+p_T^2}$ is the transverse energy of charmonium. The density of gluons $f_g$ is described by a Bose distribution. The step function $\Theta(T - T_c)$ ensures that gluon dissociation occurs only in deconfined matter, which is above the critical temperature $T_c = 0.165$ GeV. The charmonium dissociation probability is given by $W_{g\psi}^{c\bar{c}} = 4\sigma_{g\psi}^{c\bar{c}}(s) F_{g\psi}(s)$~\cite{Zhu:2004nw}, where $s$ is the center-of-mass energy of the gluon and the charmonium, $F_{g\psi}$ is the flux factor, and $\sigma_{g\psi}^{c\bar{c}}$ is the gluon-dissociation cross-section, calculated using the method of Operator Product Expansion~\cite{Peskin:1979va,Bhanot:1979vb},
\begin{align}
\sigma_{gJ/\psi}=A_0 {(x-1)^{3/2}\over x^5}.
\end{align}
Here $x$ is the ratio of the gluon energy to the in-medium binding energy $\epsilon_\psi$ of $J/\psi$, which incorporates modifications from the color screening effect. The constant factor $A_0 = \frac{2^{11} \pi}{27 \sqrt{m_c^3 \epsilon_\psi}}$ uses the mass of the charm quark $m_c = 1.87$ GeV. The dissociation rates for the excited states ($\chi_c, \psi'$) are determined by the geometry scale~\cite{Chen:2018kfo}.

The regeneration rate of charmonium depends on the densities of charm and anti-charm quarks and their recombination probability,
\begin{align}
{\beta_\psi}(\boldsymbol{p},{\bf x}) &= \frac{1}{2E_T}\int\frac{d^3\boldsymbol{p}_g}{(2\pi)^3 2E_g}\frac{d^3\boldsymbol{p}_c}{(2\pi)^3 2E_c}\frac{d^3\boldsymbol{p}_{\bar{c}}}{(2\pi)^3 2E_{\bar{c}}}  \notag
\\ &\times W_{c\bar{c}}^{g\psi}(s)f_c(\boldsymbol{p}_c,{\bf x})f_{\bar{c}}(\boldsymbol{p}_{\bar{c}},{\bf x}) \notag
\\&\times \Theta(T(\boldsymbol{x})-T_c)(2\pi)^4\delta(p+p_g-p_c-p_{\bar{c}}), \label{regeneration}
\end{align}
where \( W_{c\bar{c}}^{g\psi} \) is the recombination probability, determined through the detailed balance with $W_{g\psi}^{c\bar{c}}$ used in the decay rate. Previous studies indicate that charm quarks are strongly coupled with the expanding QGP. The large mass of charm quarks delays the time of their kinetic thermalization within the short lifetime of QGP expansion~\cite{He:2021zej,Chen:2016mhl,Du:2022uvj}. The degree of kinetic thermalization of the charm quarks in the reaction of charmonium regeneration depends both on the local temperatures of the QGP and the initial momentum of the charm quarks~\cite{Yang:2023rgb}. The effect of nonthermal momentum distributions of charm quarks on the production and collective flows of regenerated charmonium has been studied in the transport model~\cite{He:2021zej,Pan:2023ouw}. It enhances the $R_{AA}(p_T)$ and $v_2(p_T)$ of charmonium in middle and high $p_T$ regions, which provides a better explanation of the experimental data. For simplicity, we assume instant kinetic thermalization of charm quarks at $\tau=\tau_0$ where the hydrodynamic equations start. This simplification may lead to a slightly underestimation of the regenerated yield in the mid-$p_T$ region. The spatial density of charm quarks is governed by the conservation equation  $\partial_\mu (\rho_c(\tau, \mathbf{x}) u^\mu) = 0 $, where  $u^\mu$ is the four-velocity of the expanding QGP as given by the hydrodynamic equations and $\rho_c(\tau, \mathbf{x})$ is the spatial density of charm quarks. The momentum distribution of charm quarks is modeled as a normalized Fermi distribution. The initial cross sections for charm pairs will be given in the next section. 

To describe the dynamical evolution of charmonium in the QGP, we also need the temperature profiles of QGP generated by the 2+1 dimensional hydrodynamic model. The viscosity correction is neglected, considering that it gives less modification to the temperature profiles of the medium compared to the viscosity effects on collective flows of light hadrons~\cite{Shen:2014vra}. For the equation of state used in the hydrodynamic model, the deconfined phase is treated as an ideal gas consisting of light quarks, strange quarks, and massless gluons. The hadronic gas consists of hadrons with a mass of up to 2 GeV~\cite{ParticleDataGroup:2020ssz}. 
In the $\sqrt{s_{NN}}=5.02$ TeV Pb-Pb collisions, the initial energy density as input to the hydrodynamic equations can be determined via the multiplicity of light-charged hadrons measured by experiments. We estimate the initial maximal temperatures of the QGP to be $T_0({\bf x}_T=0,\tau_0|b=0)=510$ MeV~\cite{Zhao:2017yhj} in the central rapidity bins in the most central collisions. While in the forward rapidity bin, the initial maximal temperature of the QGP is extracted to be 450 MeV~\cite{Chen:2018kfo}. $\tau_0=0.6$ fm/c is the time of the local equilibrium of the medium and the beginning of the hydrodynamic equations~\cite{Shen:2014vra}. The initial temperatures at other positions and collision centralities can be easily obtained via the Glauber model.

\section{Initial conditions}

\begin{figure}[htbp!]
\centering
\includegraphics[width=0.37\textwidth]{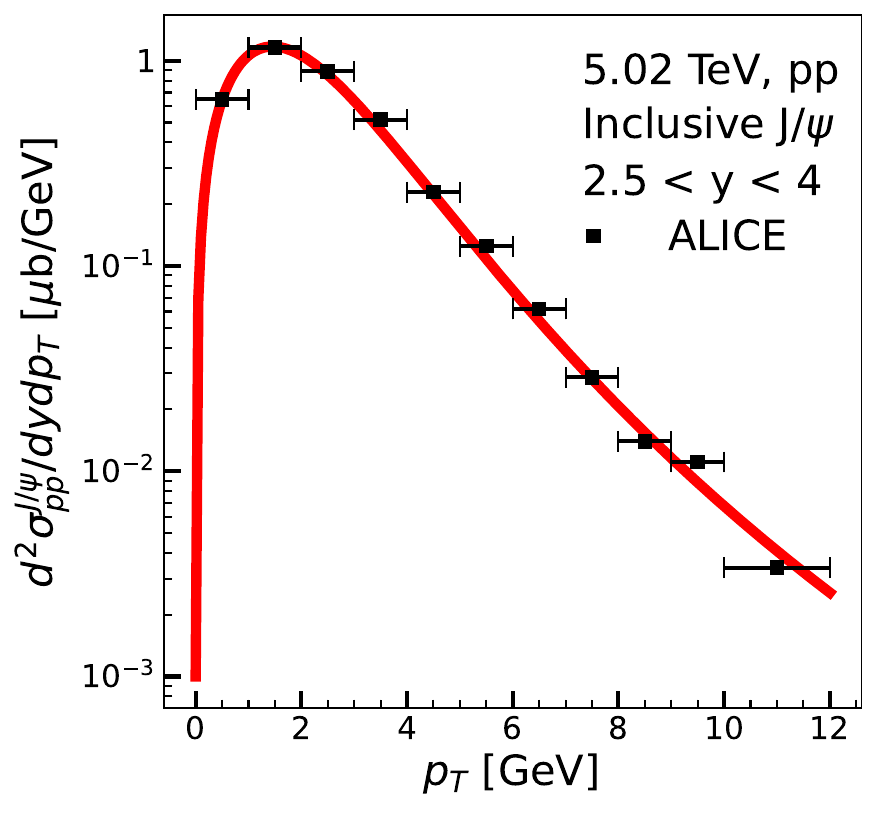}
\caption{ Differential cross-section of inclusive $J/\psi $ as a function of transverse momentum $ p_T$ in the forward rapidity region for   $\sqrt{s_{NN}} = 5.02$  TeV pp collisions. The experimental data are from the ALICE Collaboration \cite{ALICE:2016flj}.}
\label{fig:dsigma_dydpt}
\end{figure}

The initial conditions of the heavy quarkonium are needed to solve the transport model. In heavy-ion collisions, the initial distribution of heavy quarkonium can be treated as a superposition of the distribution in the nucleon-nucleon collisions, with the modification from cold nuclear matter effects. The initial spatial distribution of charmonium is proportional to the number of binary collisions $n_{coll}({\bf x}_T)$. The initial momentum distribution of charmonium has been measured in pp collisions. In pp collisions at $\sqrt{s_{NN}} = 5.02$ TeV, the initial differential cross-section of inclusive $J/\psi$ is parameterized with the form~\cite{Zhou:2014kka},
\begin{align}
&\frac{d^2\sigma_{pp}^{J/\psi}}{dy2\pi p_Tdp_T}= \nonumber \\
&\frac{2(n-1)}{2\pi(n-2)\langle p_T^2\rangle_{pp}^{J/\psi}}[1+\frac{p_T^2}{(n-2)\langle p_T^2\rangle_{pp}^{J/\psi}}]^{-n}  
\frac{d\sigma_{pp}^{J/\psi}}{dy}, \label{eq.ppinit}
\end{align}
where  $y$  denotes the rapidity of  $J/\psi$. The parameters $\langle p_T^2 \rangle_{pp}^{J/\psi}$  and $ n $ control the initial transverse momentum distribution of $J/\psi$. To incorporate the rapidity dependence in the transverse momentum distribution, $\langle p_T^2 \rangle_{pp}^{J/\psi}$  is parametrized as 
$
\langle p_T^2 \rangle_{pp}^{J/\psi} = \langle p_T^2 \rangle_{y=0} \times (1 -  {y^2}/{Y_\psi^2}),
$
where $Y_\psi \equiv \text{arccosh} \left[{\sqrt{s_{NN}}}/(2m_\psi)\right]$ ~\cite{Chen:2015iga} represents the maximum rapidity of $J/\psi$ with zero transverse momentum, and $m_\psi$ is the mass of $J/\psi$. Given that experiments have measured the $J/\psi$ differential cross section in the forward rapidity bin $ 2.5 < y < 4 $, we integrate Eq. (\ref{eq.ppinit}) over this forward rapidity bin to extract the values   $\langle p_T^2 \rangle_{y=0} = 10.6\ \text{(GeV/c)}^2$  and  $n = 3.5$ for 5.02 TeV pp collisions. The corresponding parametrization line and the experimental data are shown in Fig. \ref{fig:dsigma_dydpt}.

\begin{figure}[htbp!]
\centering
\includegraphics[width=0.35\textwidth]{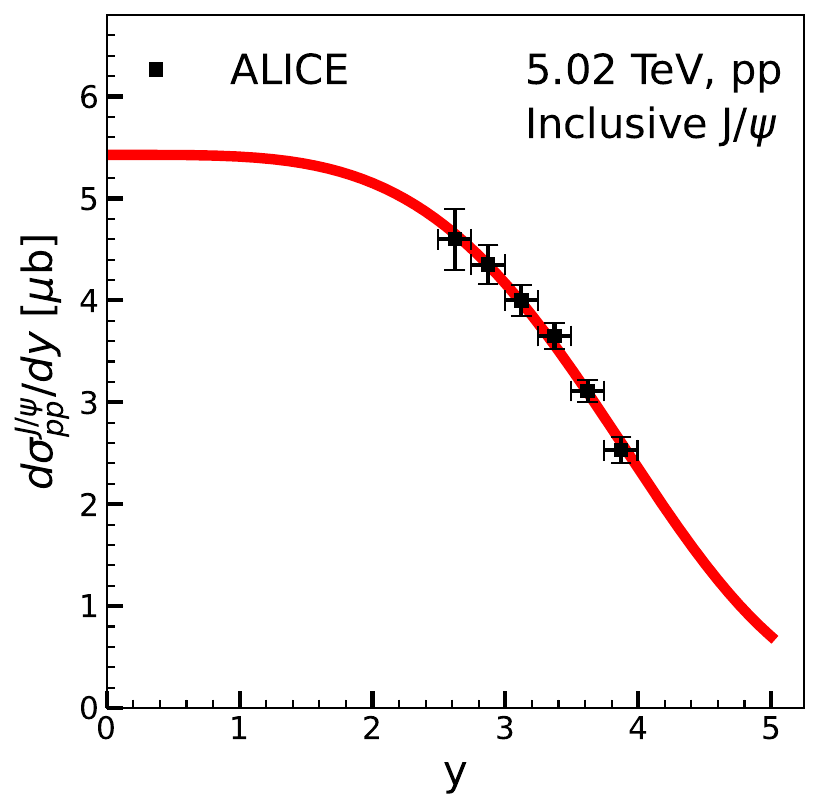}
\caption{ $J/\psi$ inclusive production cross section as a function of rapidity $y$ in pp collisions at 5.02 TeV. Experimental data is from ALICE Collaboration \cite{ALICE:2017leg}.}
\label{fig:dsigma/dy_pp}
\end{figure}

In heavy-ion collisions, the initial mean transverse momentum square of $J/\psi$ is enhanced compared to proton-proton collisions, called the Cronin effect~\cite{Cronin:1974zm}. We include this effect by modifying $\langle p_T^2\rangle_{pp}^{J/\psi}$ with $\langle p_T^2\rangle^{J/\psi}=\langle p_T^2\rangle_{pp}^{J/\psi}+a_{gN}\cdot \langle L\rangle$, where the parameter $a_{gN}=0.15\ \rm{GeV^2/fm}$~\cite{Zhou:2014kka} represents the mean energy squared per unit length that partons obtained via interacting with other nucleons before generating charmonium. $\langle L\rangle$ is the mean length through which partons travel before charmonium production, which can be calculated with the nuclear thickness function. 
The rapidity dependence in the cross section is introduced in the parametrization~\cite{Chen:2015iga}, 
\begin{align}
{d\sigma_{pp}^{J/\psi} \over dy}= A \exp{\left(-\frac{y^4}{2B^2}\right)},
\end{align}
where 
the parameters $A $ and $B $ are determined with the ALICE data~\cite{ALICE:2017leg} to be $ A = 5.43\ \mu\text{b}$ and $ B = 12.4 $. The fitted line is compared with the experimental data in Fig.\ref{fig:dsigma/dy_pp}. 
The ratio of the prompt over inclusive $J/\psi$ production cross-sections is $1-f_B(p_T)$, where $f_B$ is fitted to be $f_B=0.04+0.023p_T/(GeV/c)$ based on the experimental data in pp collisions at different collision energies~\cite{CDF:2004jtw,CMS:2010nis}.
 For the charm pair production cross-section, its rapidity dependence is assumed to be the same as that of  $J/\psi $,  $d\sigma_{pp}^{c\bar{c}}/dy = 220 \times d\sigma_{pp}^{J/\psi}/dy  $, where the ratio of 220 is determined from experimental data on the production cross-sections of  $J/\psi $ and $c\bar{c}$ pair at 5.02 TeV pp collisions~\cite{Chen:2015iga}. In nuclear collisions, the parton density of nucleons is modified by surrounding nucleons compared to the case in proton-proton collisions. Consequently, the production cross-sections of the $J/\psi$ and $ c\bar{c} $  pairs are also adjusted due to the shadowing effect~\cite{Mueller:1985wy}. This modification factor varies with the collision energy and is calculated using the EPS09 package~\cite{Eskola:2009uj}. The shadowing factor exhibits a clear dependence on rapidity, showing distinct variations in forward and backward rapidities. This behavior is particularly evident in p-Pb collisions~\cite{ALICE:2013snh}. In the backward rapidity region, where the effects of the hot medium are more pronounced, the nuclear modification factor of $J/\psi$ is enhanced due to the anti-shadowing effect from the Pb nucleus~\cite{Ferreiro:2014bia,Chen:2016dke}. In contrast, for Pb-Pb collisions, the combined shadowing factor, resulting from the product of the shadowing effects of both nuclei, is nearly symmetric in both forward and backward rapidities, with a reduced variation across rapidities~\cite{ALICE:2012jsl,Chen:2015iga}. Therefore, in the following calculations, the shadowing factor for \( J/\psi \) and \( c\bar{c} \) is taken as 0.8 and 1.0 respectivley in forward rapidities in the \( \sqrt{s_{NN}} = 5.02 \) TeV Pb-Pb collisions~\cite{Chen:2015iga}.

 The shadowing effect on bottom quark pairs is relatively weak due to the large mass of bottom quarks, and this effect is neglected in the calculation of the non-prompt part of the $J/\psi$ nuclear modification factor. Meanwhile, in peripheral collisions, a small hot medium is generated in the overlap region of the two nuclei, while coherent and incoherent photoproduced charmonium are distributed over the nuclear area. This suggests that only a small fraction of the photoproduced charmonium resides within the QGP. The modification of photoproduction due to the hot medium is minimal and is therefore neglected in this study. The effects of cold nuclear matter are also not incorporated in photoproduction, which will be studied in detail in the following works. 

\section{Numerical results in heavy-ion collisions}

\begin{figure}[htbp!]
\centering
\includegraphics[width=0.33\textwidth]{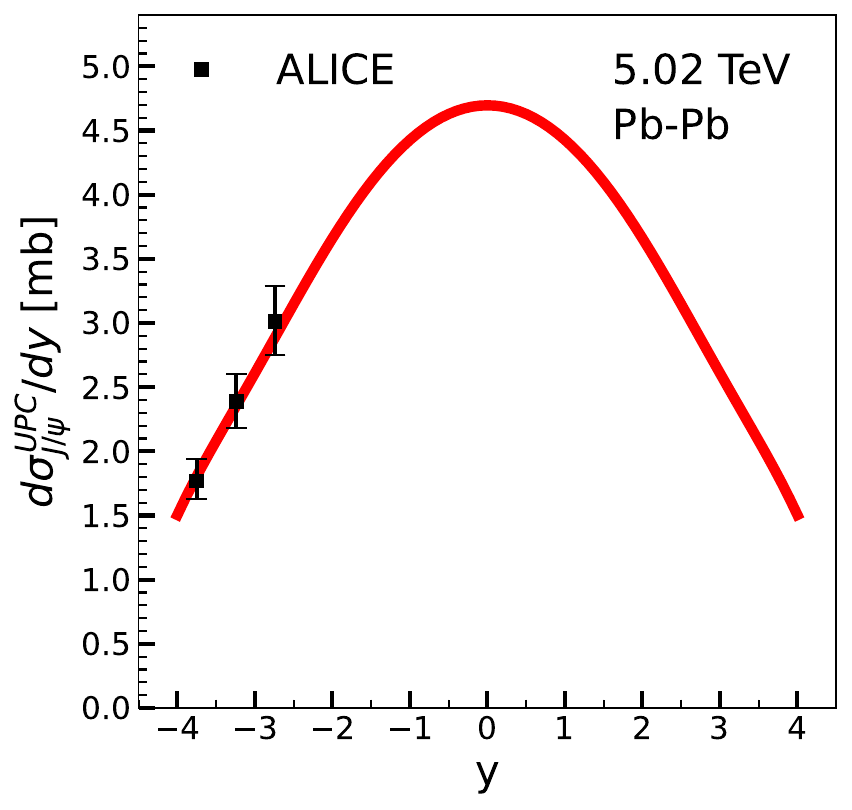}
\includegraphics[width=0.33\textwidth]{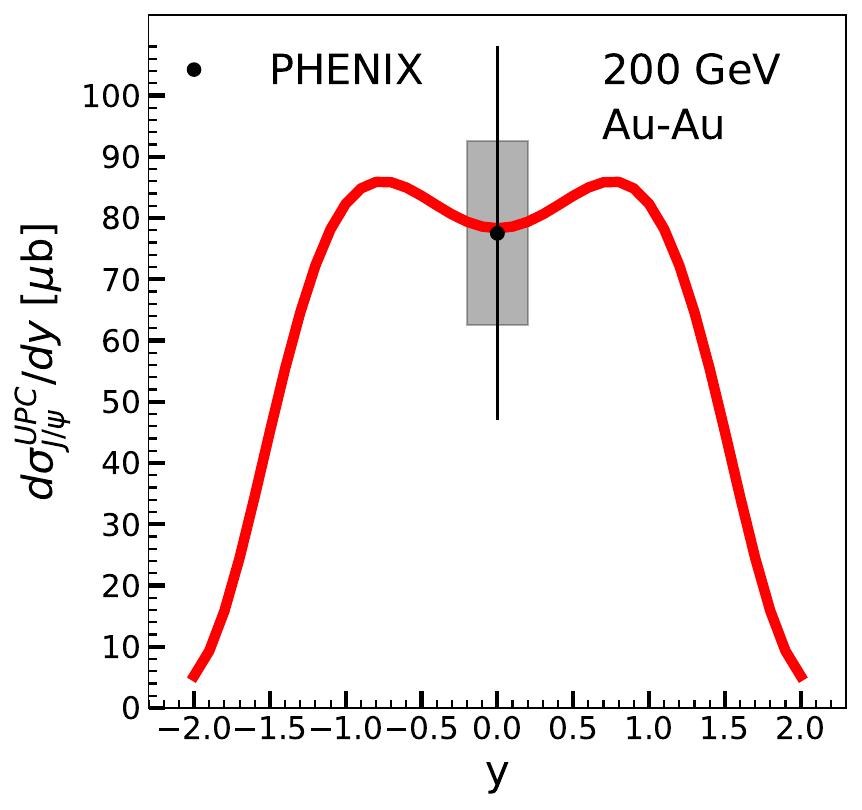}
\caption{ The rapidity differential cross section of $J/\psi$ photoproduction in UPC at 5.02 TeV Pb-Pb collisions (upper panel) and 200 GeV Au-Au collisions (lower panel). The experimental data are cited from ALICE Collaboration~\cite{Pozdniakov:2017fwn} and PHENIX Collaboration\cite{PHENIX:2009xtn}.}
\label{fig:UPC}
\end{figure}

In ultra-peripheral collisions, the production of $J/\psi$ is dominated by coherent photoproduction. We apply the EPA method outlined in Section II to calculate the coherent photoproduction cross-section of $J/\psi$ in ultra-peripheral collisions of 5.02 TeV Pb-Pb and 200 GeV Au-Au collisions, as shown in Fig.\ref{fig:UPC}. For Au-Au collisions, the nuclear thickness function has been updated, and the parameters for the Pomeron exchange process have been revised to \( X = 0.0015\ \mu\text{b} \) and \( \epsilon = 0.68 \)~\cite{Klein:1999qj}. The rapidity-differential cross-sections, shown in Fig.\ref{fig:UPC}, are calculated and demonstrate good agreement with experimental data for $J/\psi$ in both central and forward rapidities at the two collision energies.

\begin{figure}[htbp!]
\centering
\includegraphics[width=0.33\textwidth]{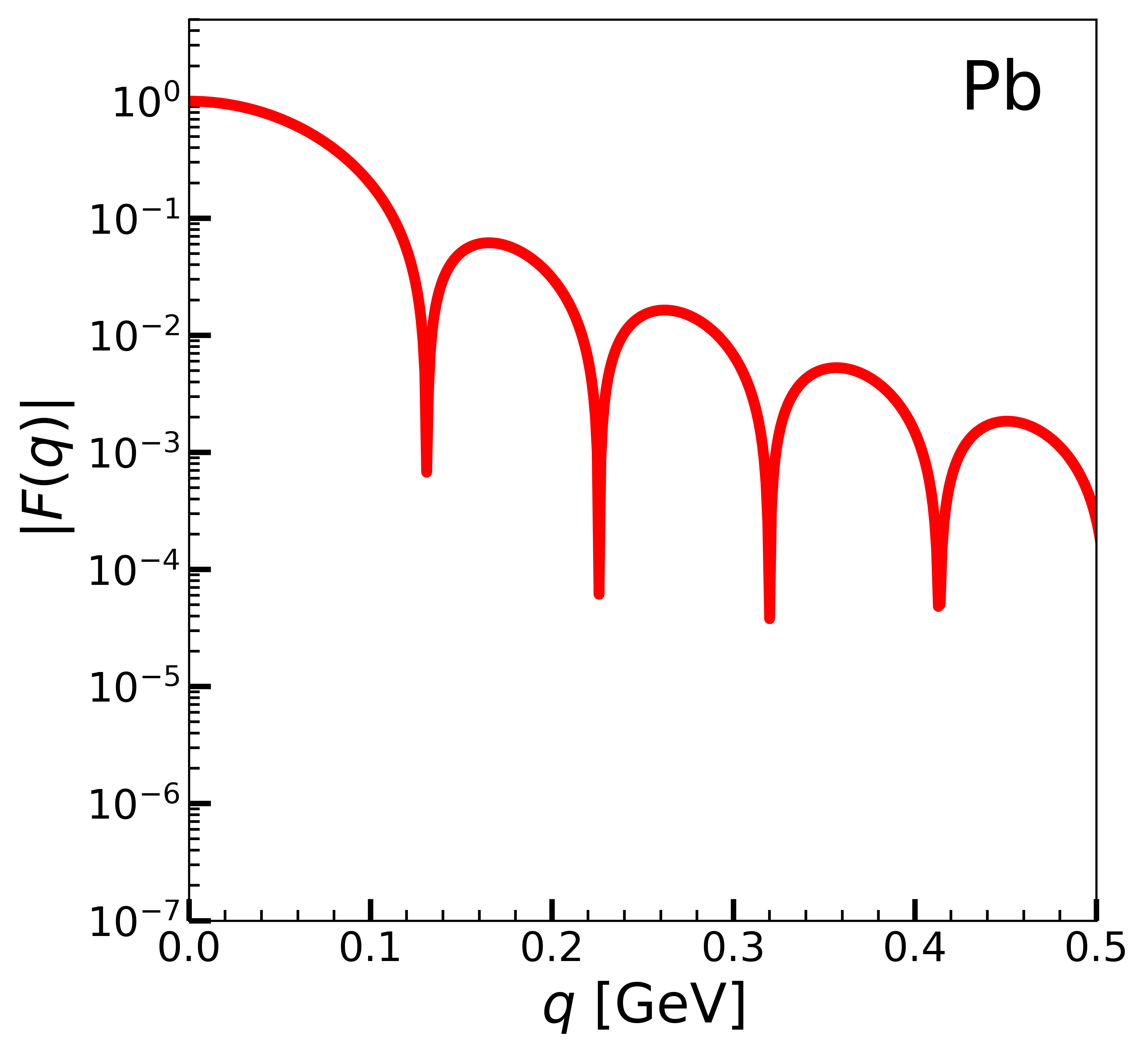}
\caption{ Form factor for the nucleus Pb as a function of momentum $q$. }
\label{fig:form_factor}
\end{figure}

In the \( p_T \) distribution of \( J/\psi \), the \( p_T \)-dependence is introduced through the form factor \( F(q) \) of the electric charge density in the nucleus, as shown in Eq. (\ref{eq.sigmaA}) and Eq. (\ref{eq.ptphotores}). We present the plot of \( |F(q)| \) for the nucleus Pb as an example in Fig. \ref{fig:form_factor}. An evident oscillation in \( |F(q)| \) is observed, resulting from the Fourier transform in Eq. (\ref{eq.form}). This oscillation will be visible in the \( p_T \) distribution of the photoproduced \( J/\psi \) and in the \( R_{AA} \) in the small \( p_T \) region. Furthermore, the value of \( |F(q)| \) decreases by one order of magnitude when \( q \) is around 0.2, suggesting that the coherent photoproduction is suppressed by two orders of magnitude at the same \( q \).

\begin{figure}[htbp!]
\centering
\includegraphics[width=0.33\textwidth]{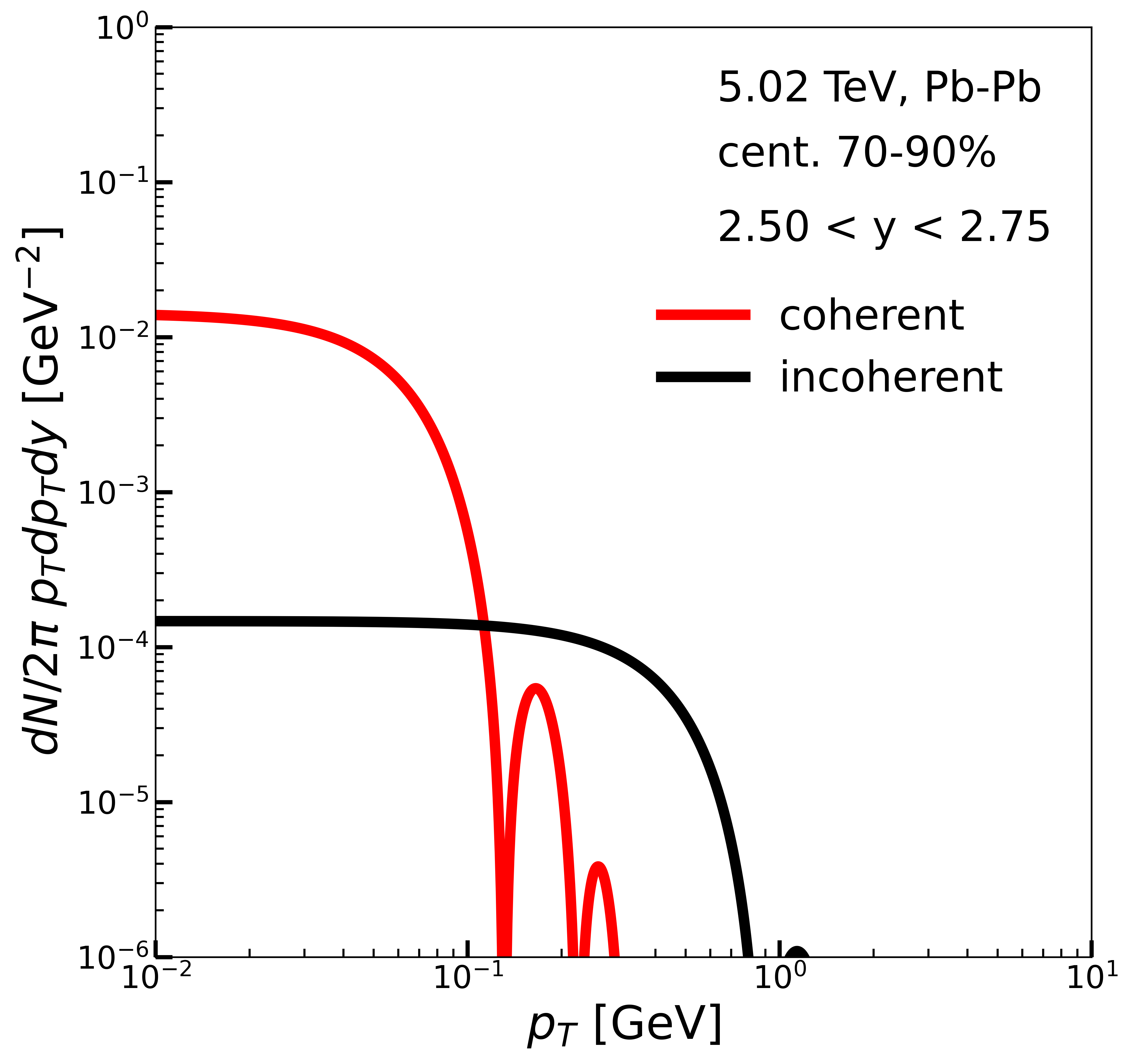}
\caption{ Comparison between coherent and incoherent photoproduction of $J/\psi$ in peripheral collisions with the centrality 70-90\% in 5.02 TeV Pb-Pb collisions. The red and black lines are for coherent and incoherent photoproduction, respectively. }
\label{fig:compare-twophoto}
\end{figure}

For incoherent photoproduction, its \( p_T \)-dependence is introduced through the form factor of nucleons (Eq. (\ref{eq:formproton})), which causes the incoherent photoproduction to be distributed over a broader range of \( p_T \). In Fig. \ref{fig:compare-twophoto}, when \( p_T \) is less than 0.1, coherent production is significantly greater than the incoherent part. However, at higher momentum values, the incoherent contribution becomes dominant. This conclusion is consistent with the previous estimation that the two mechanisms dominate in the \( p_T \) regions \( p_T < 1/R_A \) and \( 1/R_A < p_T < 1/R_N \), respectively.

Finally, we present the inclusive nuclear modification factors $R_{AA}$, which account for all contributions, including photoproduction, prompt production (primordial production + regeneration), and non-prompt production from B-hadron decay \cite{Chen:2013wmr}, 
\begin{align}
\label{eq-inclusive}
R_{AA}&={N_{AA}^{\rm init+rege}+N_{AA}^{ B\rightarrow J/\psi}+N_{AA}^{\rm photo}\over N_{pp}^{J/\psi}N_{coll}}, 
\end{align}
where $N_{AA}^{\rm init+rege}$ is the sum of primordial production and regeneration given by the Boltzmann transport equation. 
\( N_{AA}^{\rm photo} \) represents the sum of coherent and incoherent photoproduction of \( J/\psi \). \( N_{pp}^{J/\psi} \) and \( N_{coll} \) denote the \( J/\psi \) production in pp collisions and the number of binary collisions, respectively. The term \( {N_{AA}^{B \rightarrow J/\psi}}/({N_{pp}^{J/\psi} N_{coll}}) \) represents the ratio of non-prompt to inclusive \( J/\psi \) production in AA collisions, incorporating the modification from bottom quark energy loss in the QGP. This is equivalent to \( f_B(p_T) \times Q(p_T) \). The hot medium effects are encoded in the quenched factor \( Q(p_T) \), which is defined to be the ratio of the final to initial momentum distributions of bottom quarks as they evolve in the QGP. This can be calculated using the Langevin model~\cite{Yang:2023rgb}. In the centrality range 70-90\%, where the effects of the hot medium are weak, the quenched factor is extracted to be 0.9 \cite{Yang:2023rgb}, with minimal momentum dependence in the relevant \( p_T \) region of this study, such as \( p_T \lesssim 3 \) GeV/c.

\begin{widetext}

\begin{figure}[htp!]
    \centering
    \begin{tikzpicture}
        \node[anchor=south west,inner sep=0] (image1) at (0,0) {\includegraphics[width=0.26\textwidth]{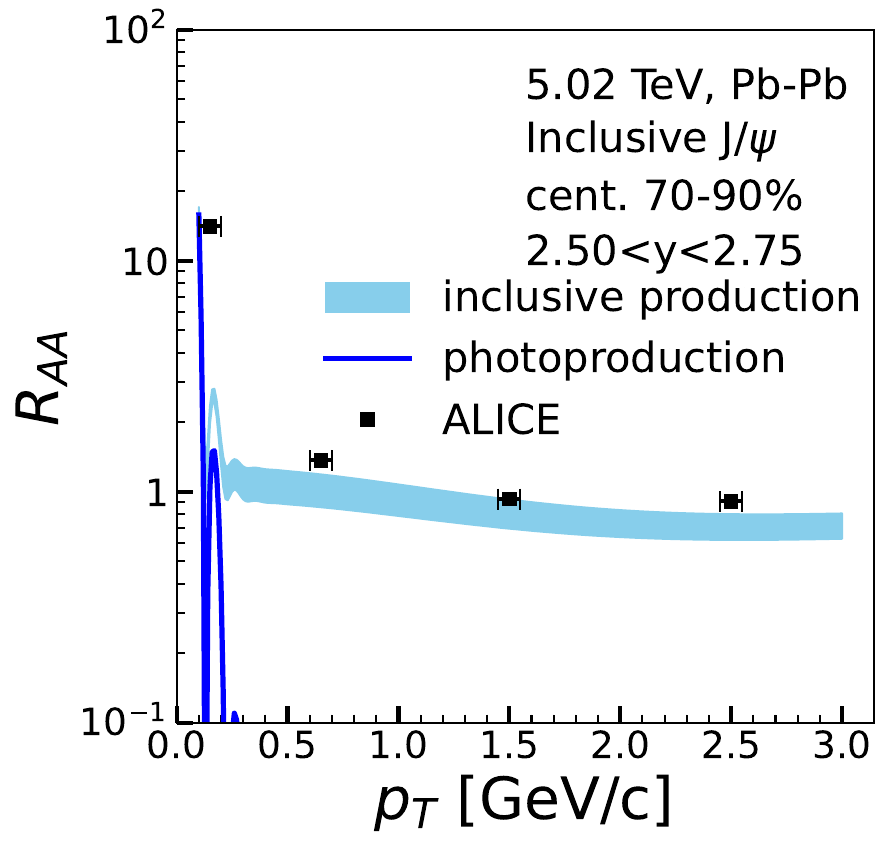}};
        \begin{scope}[x={(image1.south east)},y={(image1.north west)}]
        \end{scope}
    \end{tikzpicture}
    \begin{tikzpicture}
        \node[anchor=south west,inner sep=0] (image2) at (0,0) {\includegraphics[width=0.26\textwidth]{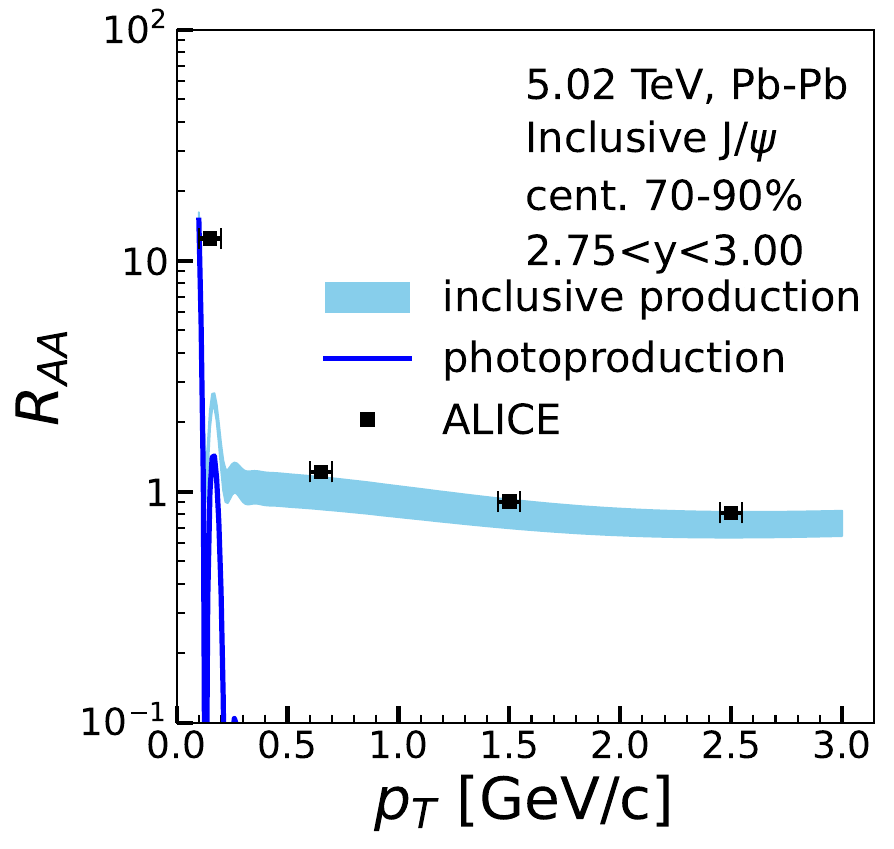}};
        \begin{scope}[x={(image2.south east)},y={(image2.north west)}]
        \end{scope}
    \end{tikzpicture}
    \begin{tikzpicture}
        \node[anchor=south west,inner sep=0] (image1) at (0,0) {\includegraphics[width=0.26\textwidth]{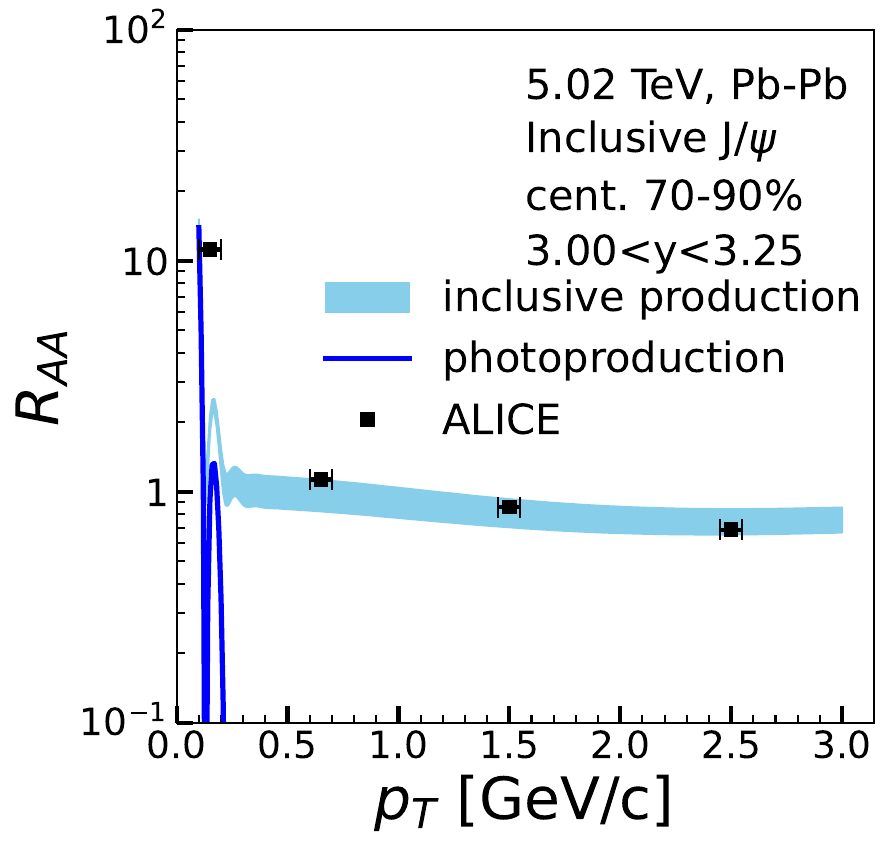}};
        \begin{scope}[x={(image1.south east)},y={(image1.north west)}]
        \end{scope}
    \end{tikzpicture}
    \begin{tikzpicture}
        \node[anchor=south west,inner sep=0] (image1) at (0,0) {\includegraphics[width=0.26\textwidth]{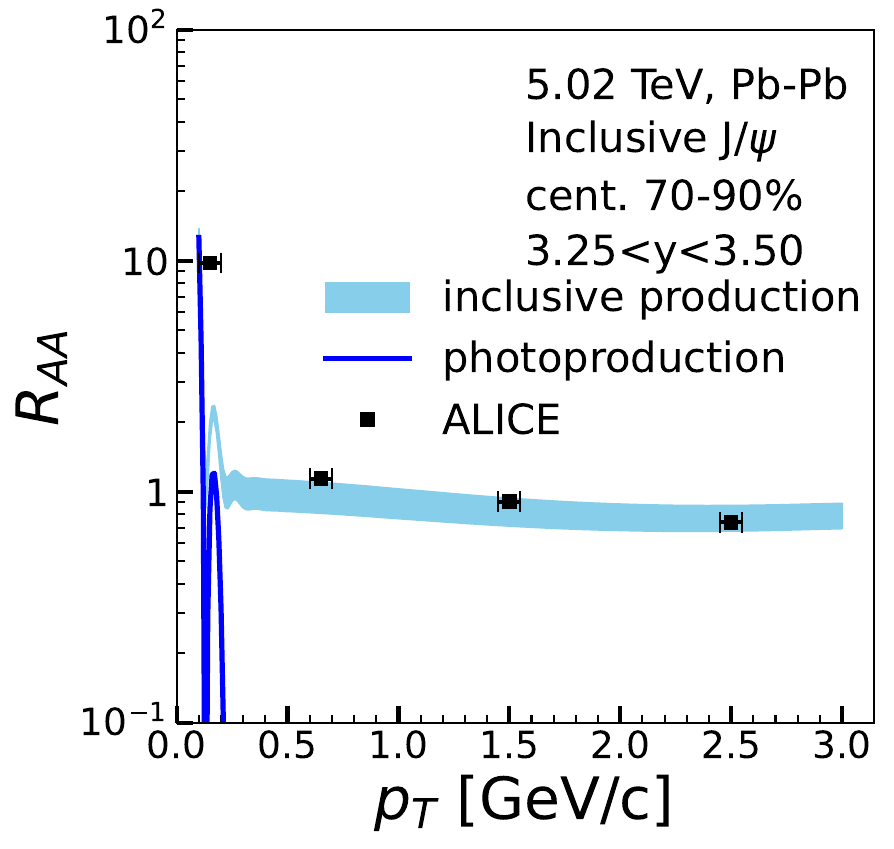}};
        \begin{scope}[x={(image1.south east)},y={(image1.north west)}]
        \end{scope}
    \end{tikzpicture}
    \begin{tikzpicture}
        \node[anchor=south west,inner sep=0] (image1) at (0,0) {\includegraphics[width=0.26\textwidth]{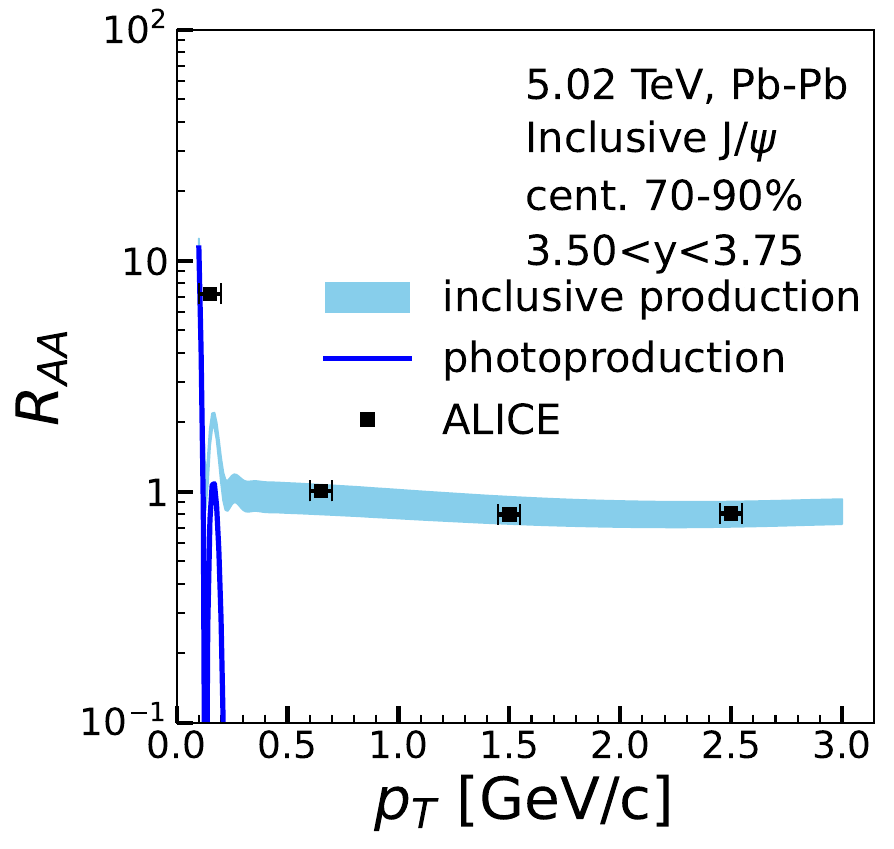}};
        \begin{scope}[x={(image1.south east)},y={(image1.north west)}]
        \end{scope}
    \end{tikzpicture}
    \begin{tikzpicture}
        \node[anchor=south west,inner sep=0] (image1) at (0,0) {\includegraphics[width=0.26\textwidth]{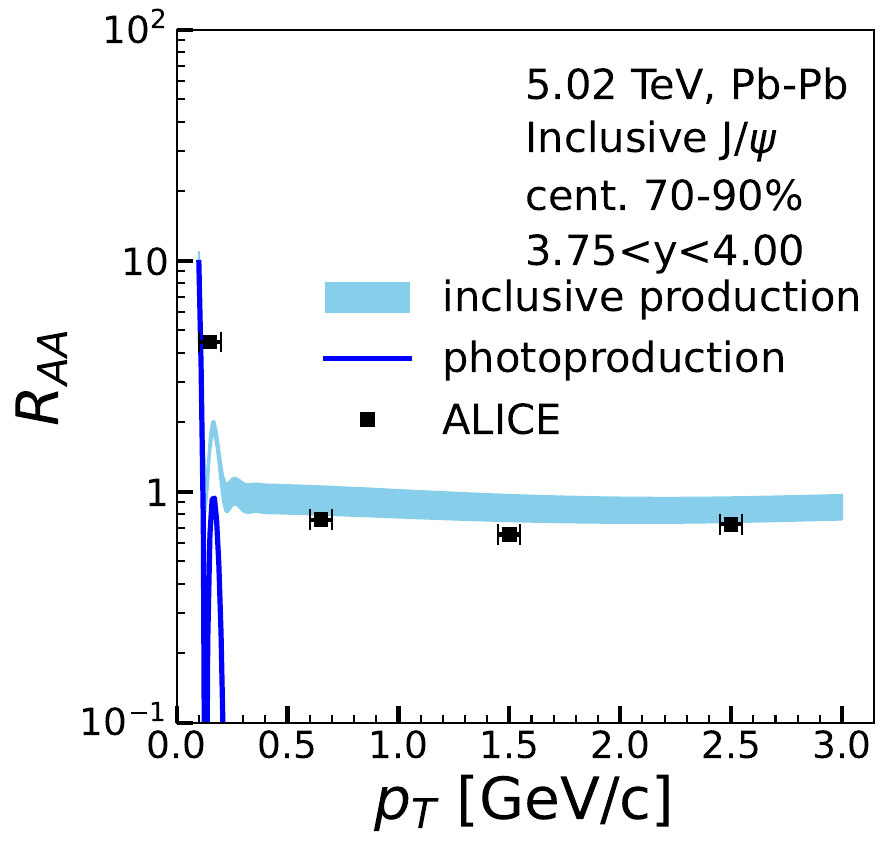}};
        \begin{scope}[x={(image1.south east)},y={(image1.north west)}]
        \end{scope}
    \end{tikzpicture}
    \caption{Inclusive nuclear modification factor $R_{AA}$ as a function of transverse momentum in the centrality 70-90\% at 5.02 TeV Pb-Pb collisions. Six rapidity bins are considered. The blue line is for photoproduction. The band represents inclusive production, including prompt production, non-prompt production from B-decay, and coherent and incoherent photoproduction. The upper and lower limits of the band correspond to the uncertainty of the shadowing factor in hadroproduction. The shadowing factor of charmonium is taken as 0.8 and 1.0, respectively.  
    The experimental data are cited from the ALICE Collaboration~\cite{Shatat:2024nqb}.}
    \label{fig:bmodel}
\end{figure}

\end{widetext}

In Fig.\ref{fig:bmodel}, the \( p_T \)-differential inclusive nuclear modification factor for \( J/\psi \) is calculated in six rapidity bins for 5.02 TeV Pb-Pb collisions. The solid blue line represents the sum of coherent and incoherent photoproduction, while the band indicates the total production, calculated using Eq.(\ref{eq-inclusive}). The lower and upper limits of the band correspond to the shadowing factor in hadroproduction, taken as 0.8 and 1.0, respectively. Since photoproduction is suppressed in pp collisions due to the small number of electric charges, the additional term \( N_{AA}^{\rm photo} \) significantly enhances the value of \( R_{AA} \) in the region \( p_T < 1/R_N \approx 0.2 \) GeV/c, where photoproduction dominates. The value of \( R_{AA} \) approaches infinity in UPC because the denominator approaches zero. 
In higher \( p_T \) bins, photoproduction becomes negligible, and hadroproduction dominates. For regeneration, it is generally important in the \( p_T \lesssim 3 \) GeV/c range~\cite{Zhou:2014kka,Zhao:2011cv}. However, regeneration is suppressed in peripheral collisions due to the small number of charm pairs, while primordial production dominates hadroproduction.

\section{Summary}

In this study, we utilize the equivalent photon approximation to explore the coherent and incoherent photoproduction of $J/\psi$ in ultraperipheral and peripheral heavy-ion collisions at RHIC and LHC energies. Using a Boltzmann-type transport model that incorporates the cold and hot nuclear matter effects, we investigate hadroproduction, consisting of $J/\psi$ from initial production, the recombination of charm and anti-charm quarks, and non-prompt production from B-decay. In peripheral collisions, photoproduction prevails at extremely low transverse momentum, such as $p_T<0.2$ GeV/c, exhibiting a significant dependency on photon densities. This enhances the value of the nuclear modification factor to be around 10. At higher $p_T$, primarily produced $J/\psi$ and non-prompt $J/\psi$ from B-decay predominate the inclusive production in peripheral collisions. At the same time, the regeneration contribution is small due to a small number of charm pairs in QGP. Our theoretical studies have included all the contributions to the production of $J/\psi$ and align well with the experimental data. This insight helps to understand the intensity of electromagnetic fields and their role in vector meson photoproduction in heavy-ion collisions.

\vspace{0.8cm}
{\bf Acknowledgment:} This work is supported by the National Natural Science Foundation of China
(NSFC) under Grant Nos. 12175165. 
\nocite{*}

\bibliography{paper}

\end{document}